\documentclass[preprint,preprintnumbers,amsmath,amssymb]{revtex4}

\usepackage{epsfig}
\usepackage{graphicx}% Include figure files
\usepackage{dcolumn}% Align table columns on decimal point
\usepackage{bm}% bold math
\usepackage{url}
\usepackage[dvipsnames]{xcolor}
\usepackage[colorinlistoftodos,prependcaption,textsize=small]{todonotes}

\begin{document}

%\documentclass{article}
%\usepackage[utf8]{inputenc}

%\nofiles

\title{Local Granger Causality}

\author{ Sebastiano Stramaglia$^{1}$, Tomas Scagliarini$^{1}$, Yuri Antonacci$^{2}$, and Luca Faes$^{3}$}

\affiliation{$^1$ Dipartimento Interateneo di Fisica, Universit\'a degli Studi di Bari Aldo Moro, and INFN, Sezione di Bari, 70126 Bari, Italy\\}

\affiliation{$^2$ Dipartimento di Fisica e Chimica, Universitá di Palermo, 90123 Palermo, Italy \\ }

\affiliation{$^3$ Dipartimento di Ingegneria, Universit\'a di Palermo, 90128, Palermo, Italy \\}

\date{\today}% It is always \today, today,

\begin{abstract}
Granger causality is a statistical notion of causal influence based on prediction via vector autoregression.  For Gaussian variables it is equivalent to transfer entropy, an information-theoretic measure of time-directed information transfer between jointly dependent processes. 
We exploit such equivalence and calculate exactly the {\it local Granger causality}, i.e. the profile of the information transfer at each discrete time point in Gaussian processes; in this frame Granger causality is the average of its local version. Our approach offers a robust and computationally fast method to follow the information transfer along the time history of linear stochastic processes, as well as of nonlinear complex systems studied in the Gaussian approximation.

%\pacs{05.10.-a,05.45.Tp,87.10.-e}
\end{abstract}\maketitle

\maketitle
Granger causality (GC) \cite{granger} and its non-parametric counterpart, transfer entropy (TE) \cite{schreiber}, are widely used tools to assess and quantify causal relationships between stochastic processes mapping the evolution of coupled dynamic systems over time. For discrete-time stationary multivariate processes represented by vector autoregressive  (VAR) models \cite{var}, GC measures the gain in the linear predictability of the target process when the knowledge of the candidate driver process is exploited to make the prediction. For Gaussian systems, GC and TE are equivalent \cite{barnett2009} and are interpreted as measures of information transfer \cite{book_te}. 

The question we address here is: is  it possible to calculate the temporal  profile of the information transfer in complex systems, so that its time-average coincides with the information-theoretic value of GC?

Concerning TE, the same question has been addressed in \cite{local_te} with the introduction of the local transfer entropy. Differently from the corresponding averaged quantity, the local transfer entropy can be both positive and negative: when it is negative at a given time step, the observation of the driver is mis-informative about the value of the target at that time. Recently, the local TE has been proposed to study phase-amplitude coupling in electrophysiological signals \cite{martinez}. In our opinion, since its inception the local transfer entropy has been used in a quite limited way w.r.t. its potentiality: the lack of non-trivial systems with an exact solution, as well as critical choices (parameters, embedding schemes) which influence the estimation of local TE, have certainly limited the popularity of this notion.
In the following we show that it is possible to calculate exactly the local Granger causality $L_{gc}$ from the parameters of the underlying VAR model. The knowledge of the exact value of the local transfer entropy in benchmark systems is helpful to get the correct interpretation of the local information transfer, in particular those of its negative values.

Let us consider a generic VAR model of order $p$ for $n$ zero-mean processes $x_1,\ldots,x_n$
\begin{displaymath}
x_i(t) = \sum_{k=1}^p\sum_{j=1}^n A_{ij}^k\, x_j(t-k) + \epsilon_i(t),
\end{displaymath}
where $A_{ij}^k$, $k=1,\ldots,p$, are $n\times n $ matrices  quantifying time lagged interactions, and the white noise innovations $\epsilon_i(t)$ have covariance matrix $\Sigma$. Under suitable conditions \cite{var} the VAR process is said to be stable and, assuming that it has been initialized in the infinite past, is stationary and ergodic with time invariant variances and covariances.
In the stationary regime, the vector ${\bf x} = [x_1(t-1)\cdots x_n(t-1) x_1(t-2) \cdots x_n(t-2) \cdots x_n(t-p)]^T$ denoting the past of the system is distributed as a multivariate Gaussian 
\begin{equation}
p({\bf x}) = \frac{\exp{-\frac{1}{2} {\bf x}^T {\bf \Psi^{-1}} {\bf x}}}{(2\pi)^{\frac{np}{2}}\sqrt{\det{\bf\Psi}}},
\label{eq:2}
\end{equation}
with covariance matrix ${\bf\Psi}$, which can be obtained from matrices ${\bf A}$ and ${\bf\Sigma}$ using Yule-Walker equations as follows. Let us introduce the $np\times np$ matrices
\[  
{\bf\hat{A}} = 
\begin{pmatrix}
{\bf A^1} & {\bf A^2} & \cdots & {\bf A^{p-1}} & {\bf A^p}   \\
{\bf I}   & {\bf 0}   & \cdots & {\bf 0}       & {\bf 0}     \\
{\bf 0}   & {\bf I}   & \cdots & {\bf 0}       & {\bf 0}     \\
\vdots  & \vdots  & \ddots & \vdots  & \vdots\\
{\bf 0} & {\bf 0} & \cdots & {\bf I} & {\bf 0}
\end{pmatrix}, \qquad
{\bf \Omega} = 
\begin{pmatrix}
{\bf \Sigma} & {\bf 0} & \cdots & {\bf 0} \\
{\bf 0}      & {\bf 0} & \cdots & {\bf 0} \\
\vdots & \vdots & \ddots & \vdots \\
{\bf 0}      & {\bf 0}      & \cdots & {\bf 0} 
\end{pmatrix}.
\]
${\bf \Psi}$ can be obtained solving the discrete time Lyapunov equation ${\bf\Psi} = {\bf\hat{A}\Psi\hat{A}^T} + {\bf\Omega}$.

Let us denote $\beta$ e $\alpha$ the indices of the driving and target variables respectively, and let $\sigma ^2$ be the variance of $\epsilon_\alpha(t)$.
Moreover, we denote $y=x_\alpha(t)$ the future state of the target,  ${\bf w} = [x_\beta(t-1)\ldots x_\beta(t-p)]^T$ the vector of driver's variables and ${\bf u}=\bf{x}\setminus {\bf w}$ the remaining variables.
Re-arranging the variables in $\bf{x}$, we write
\begin{equation*}
{\bf x} = {{\bf u} \choose {\bf w} } , \qquad 
{\bf\Psi^{-1}}= 
\begin{pmatrix}
{\bf U_0} & {\bf Z}  \\
{\bf Z^T} & {\bf U_1}
\end{pmatrix},
\end{equation*}
where the concentration matrix ${\bf \Psi^{-1}}$ has been correspondingly decomposed in blocks.

The rows $A^k(\alpha,:)$ of couplings to the target are analogously re-arranged in two vectors: $\bf{a_u}$ and $\bf{a_w}$.
We have the conditional probability distribution:
\begin{equation}
p(y|{\bf u},{\bf w})= \frac{\exp{-{{(y-{\bf a_u^Tu}-{\bf a_w^Tw})^2}\over 2\sigma^2}}}{ \sqrt{2\pi\sigma^2}},
\label{eq:1}
\end{equation}
and the joint probability distribution reads
%Possiamo quindi riscrivere $p(X)=p(u,w)$ come 
%$$
%p(u,w)= \frac{1}{(2\pi)^{\frac{np}{2}}\sqrt{det\Psi}}\me^{-\frac{1}{2}(u^T U_0u + 2 u^TZw + w^T U_1 w)} 
%$$
%Inoltre
%$$
%p(u,w,y)= \frac{1}{(2\pi)^{\frac{np}{2}}\sqrt{det\Psi}}\me^{-\frac{1}{2}(u^T U_0u + 2 u^TZw + w^T U_1 w)} \me ^{-\frac{1}{2\sigma^2}(y-A^uu-A^ww)^2}.
%$$
%Per trovare $p(u,y)$ dobbiamo marginalizzare la precedente espressione; a tale scopo, la riscriviamo come 
\begin{equation}
p(y,{\bf u},{\bf w})= {\exp{-\frac{1}{2}\left({\bf u^T} U_0{\bf u} + 2 {\bf u^T Z  w} + {\bf w^T  U_1  w} +{(y-{\bf a_u^Tu}-{\bf a_w^Tw})^2\over \sigma^2}\right)}\over {(2\pi)^{\frac{np}{2}}\sqrt{\det{\bf \Psi}}\sqrt{2\pi\sigma^2}}}.
\label{eq:3}
\end{equation}
Now we introduce the following quantities:
\begin{equation}
{\bf B}={\bf U_1} +\frac{1}{\sigma^2}{\bf a_w a_w^T}, \;\;\;\;\;\;\;\;\;
{\bf z} = \frac{1}{\sigma^2}(y-{\bf a_u^Tu}){\bf a_w} - {\bf Z^T} {\bf u},
\label{eq:4}
\end{equation}
and, integrating over $\bf{w}$, we obtain the marginal probability:
\begin{equation}
p(y,{\bf u}) = {\exp{-\frac{1}{2}\left( {\bf u^T U_0 u} +{(y-{\bf a_u^Tu})^2\over \sigma^2} -{\bf z^T B^{-1} z}\right)}\over (2\pi)^{\frac{(n-1)p}{2}}\sqrt{\det{\bf\Psi}} \sqrt{\det {\bf B}}\sqrt{ 2\pi\sigma^2}} 
\label{eq:5}
\end{equation} 
We can obtain also $p(\bf{u})$ from p({\bf u},{\bf w}):
\begin{equation}
p({\bf u}) = {\exp{-\frac{1}{2} {\bf u^T D u}} \over (2\pi)^{\frac{(n-1)p}{2}}\sqrt{\det{\bf\Psi}}\sqrt{ \det {\bf U_1}}} 
\label{eq:6}
\end{equation}
where ${\bf D}={\bf U_0}-{\bf ZU_1^{-1}Z^T}$.
The local Granger causality is twice the local transfer entropy \cite{barnett2009}:
\begin{equation}
L_{gc}({\bf u},{\bf w},y) = 2 \log\frac{p(y|{\bf u},{\bf w})p({\bf u})}{p(y,{\bf u})},\\
\label{eq:7}
\end{equation}
which reads
$$
L_{gc}({\bf u},{\bf w},y) =\log{\det{\bf B} \over \det {\bf U_1}}  +\frac{(y-{\bf a_u^Tu})^2-(y-{\bf a_u^Tu}-{\bf a_w^Tw})^2} {\sigma^2} +{\bf u^T ZU_1^{-1}Z^T u} - {\bf z^T B^{-1} z}.
$$

Note that the first term (constant w.r.t. $\bf{u}$,$\bf{w}$ and $y$) coincides with the standard definition of GC, and the remaning terms have vanishing expected value. The local Granger causality at time $t$ is thus given by $L_{gc}(t)=L_{gc}({\bf u}_t,{\bf w}_t,y_t)$, and satisfies $\langle L_{gc}(t)\rangle =GC.$

In order to characterize negative values and temporal profiles of the local Granger causality, we consider the following simple toy model:
\begin{equation}
    \begin{cases}
    y_t = \tilde{\epsilon}_t \\
   x_t = 0.2 x_{t-1} + 0.4 y_{t-1} +\epsilon_t
    \end{cases}
    \label{eq:8}
\end{equation}
where $\tilde{\epsilon}$ and $\epsilon$ are white noise terms with standard deviation $\sigma_{\tilde{\epsilon}} =1$ and $\sigma_\epsilon =0.8$.
The GC $y\to x$ is 0.18 in this case, corresponding to the mean of the local quantity $L_{gc}(t)$. In figure 1 we depict the distribution of sample points in the plane  ($\epsilon_t y_{t-1}$)-$L_{gc}(t)$, obtained from a run of eqs. (8) with length $30\times 10^6$ time steps. The plot shows that the local causality oscillates between positive and negative values, attaining large negative values when $\epsilon_t y_{t-1}$ is large and negative. The latter situation occurs when the noise pulls the system in the opposite direction w.r.t. the action of the cause $y_{t-1}$: in this case the knowledge of  $y_{t-1}$ is mis-informative about $x_t$, meaning that a reduced model implemented without using the driver performs better than the full model in (8). %(whose prediction is $0.2 x_{t-1}$) performs better  than the full model (whose prediction is $0.2 x_{t-1} + 0.4 y_{t-1}$).
Conversely, large positive values correspond to times $t$ with the noise term $\epsilon_t$ pulling the system in the same direction as the cause $y_{t-1}$. It is then worth stressing that fluctuations of $L_{gc}$ do not merely reproduce modulations of the noise of the system, but rather represent the interplay between noise and the driving variable.
These fluctuations constitute, in addition to their mean value, a hallmark of information transfer, as it can be seen in Fig. 2 where the $L_{gc}$ $y \rightarrow x$ is reported for different runs of the simulation performed changing the variance of the driving variable $y$: we find that not only the mean, but also the amplitude of the oscillations of the local GC is modulated by the strength $\sigma_{\tilde{\epsilon}}$ of the driving variable. In the limit  $\sigma_{\tilde{\epsilon}}\to 0$, we have $GC=\langle L_{gc} \rangle \sim \sigma_{\tilde{\epsilon}}^2$ and $\langle L_{gc}^2 \rangle-\langle L_{gc}\rangle^2  \sim \sigma_{\tilde{\epsilon}}^2$.

As a first application example, we take the bivariate time series of respiration (R) and heart rate (H) amplitudes measured with a sampling rate of 2 Hz from a subject suffering from sleep apneas and previously analyzed with transfer entropy \cite{schreiber} and nonlinear GC \cite{kgc}. Figure (3) shows that consecutive apneas are characterized by absence of respiratory oscillations and progressively increasing heart rate. Adopting the Gaussian approximation, these time series are fitted with a bivariate AR model of order $4$, identified with the Akaike Information criterion \cite{Akaike}. Then, we compute both the global and local GC along the two directions of interaction, as well as their significance thresholds based on iterative amplitude-adjusted Fourier Transform (IAAFT) surrogates \cite{schreiberIAAFT}. The GC is statistically significant along the direction from respiration to heart rate ($GC_{R \rightarrow H}=0.0341$, IAAFT 95$^{th}$ percentile = 0.0096), while it is low and non-significant along the opposite direction ($GC_{H \rightarrow R}=0.0015$, IAAFT 95$^{th}$ percentile = 0.0079); physiologically, this result supports the mainly unidirectional nature of respiratory sinus arrhythmia \cite{RSA}. Computation of the local GC supports the lack of interactions from heart rate to respiration, and reveals the local nature of the information transfer from respiration to heart rate: the $L_{gc}$ $R \rightarrow H$ exhibits clear marked oscillations with statistically significant mean and standard deviation only while the patient is breathing, while it is very small and non-significant during the apneas.

As another real example we consider intracranial EEG recordings from a drug-resistant epilepsy patient with an implanted array of 8 × 8 cortical electrodes and two depth electrodes with six contacts each \cite{kramer}, available at \cite{eeg}. Many studies of transfer entropy in the epileptic brain are published, see e.g. \cite{staniek}; here we analyze these signals in the Gaussian approximation. Data are sampled at 400Hz and we apply the proposed method on the fourth seizure, considering two 10-sec windows before and during the seizure. A previous paper \cite{entropy} showed that the depth electrode number 76 is close to the Seizure Onset Zone, therefore we evaluate in a pairwise fashion the local Granger causality from the depth electrode 76 to all 64 cortical electrodes. Data are fitted with a VAR with order $p$ chosen according to Akaike's criterion \cite{Akaike}: our results, averaged over the cortical targets, are displayed in figure 4. In the pre-ictal stage averaging over the targets leads to an homogenous pattern, with GC equal to 0.32; on the other hand in the ictal stage after averaging,  GC is lower (0.23) but the signal is more intermittent and shows peaks of $L_{gc}$ in correspondance of time instants in which the source coherently transmits information to a large portion of the cortical electrodes. These results show, on one side, that as expected the pattern of the information flow in the epileptic brain is different before and during the seizure. On the other hand it clearly suggests that the classical measure of Granger causality (the mean of $L_{gc}$) is not sufficient to properly describe the temporal properties of the information transfer in this system; for example, contrary to the mean $L_{gc}$, the standard deviation of $L_{gc}$ increases from 0.36 in the pre-ictal stage to 0.89. The standard deviation of $L_{gc}$ therefore conveys a description of the information transfer pattern complementary to that provided by the GC.

Summarizing, we have derived the local Granger causality for a generic VAR model. As fitting a VAR model to data only requires the choice of the order $p$, our formalism can be easily used to extract the temporal profile of information transfer for linear systems as well as for generic systems in the Gaussian approximation. We remark that in many applications nonlinearities can be neglected and the Gaussian approximation fully captures the underlying phenomena; if this is not the case, the results obtained in the Gaussian approximation still constitute the reference to which one should refer the role of nonlinearities.
We have shown that fluctuations of $L_{gc}$ are connected to the interplay between the innovation (noise) and driver processes, and that large negative (positive) values correspond to the noise pulling the system in the opposite (same) direction as the driver. Given that innovations model the environment acting on the system under consideration, and in agreement with the discussion in \cite{book_te}, we conclude that negative (mis-informative) values of $L_{gc}$ are important as they are the signature of extra features in the dynamics that are not accounted for in the past of the measured variables alone.
As GC has gained increasing popularity in many fields of science, we expect that the proposed approach will have a large impact as it allows to estimate easily the information transfer during the time history of a complex system.

%\begin{figure}[!ht] \centering
%\includegraphics[width=10cm]{image2.eps}
%\end{figure}

\begin{acknowledgments}
This research was supported by MIUR project PRIN 2017WZFTZP ``Stochastic forecasting in complex~systems''
\end{acknowledgments}

\begin{figure}[!ht] \centering
\includegraphics[width=15cm]{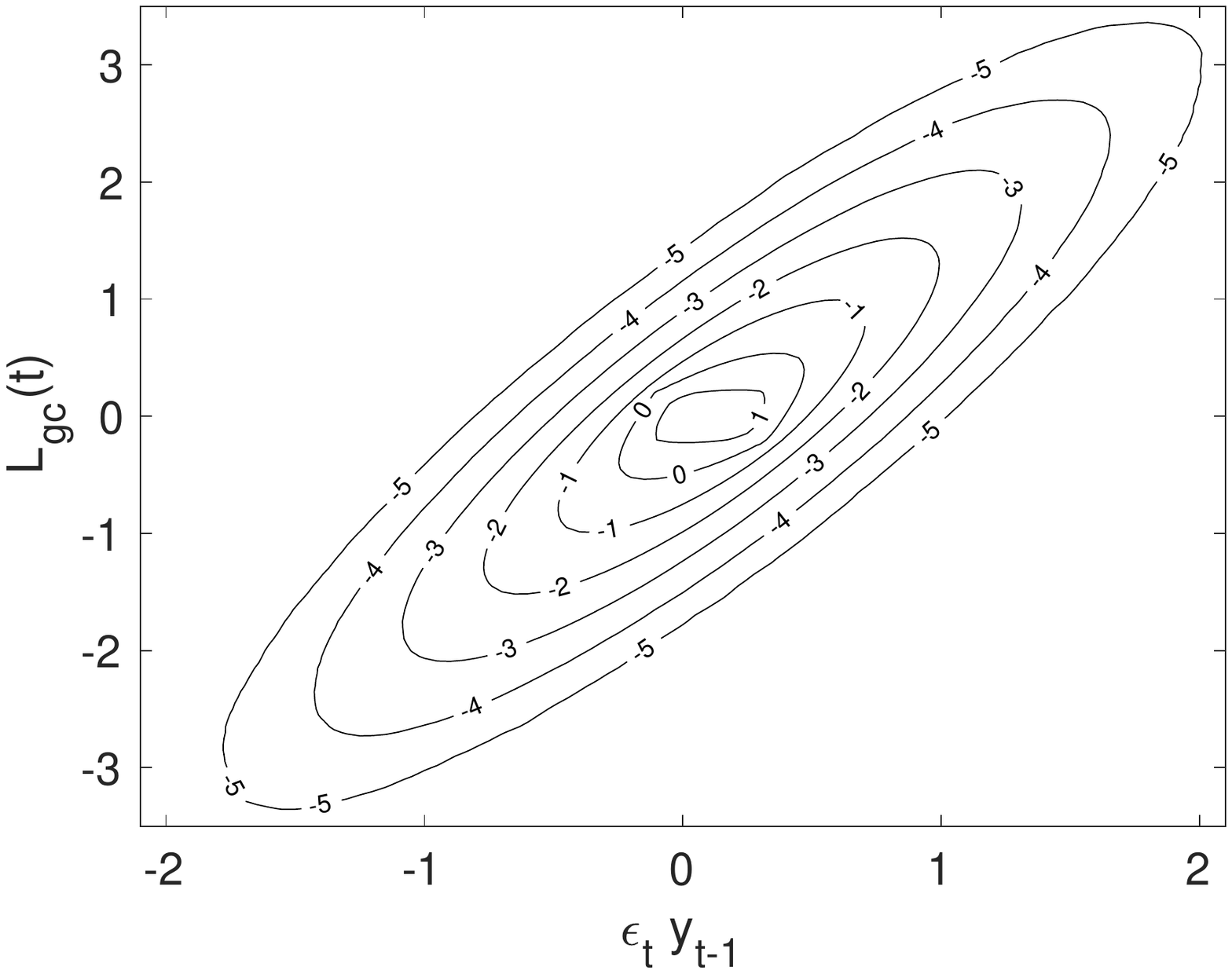}
\label{fig1}
\caption{Distribution density of the pairs ($\epsilon_t y_{t-1}$,$L_{gc}(t)$) sampled from the time evolution of model (8) depicted as a contour plot. Values displayed on the level curves are the logarihtm  of the corresponding value of the distribution density.}
\end{figure}

\begin{figure}[!ht] \centering
\includegraphics[width=15cm]{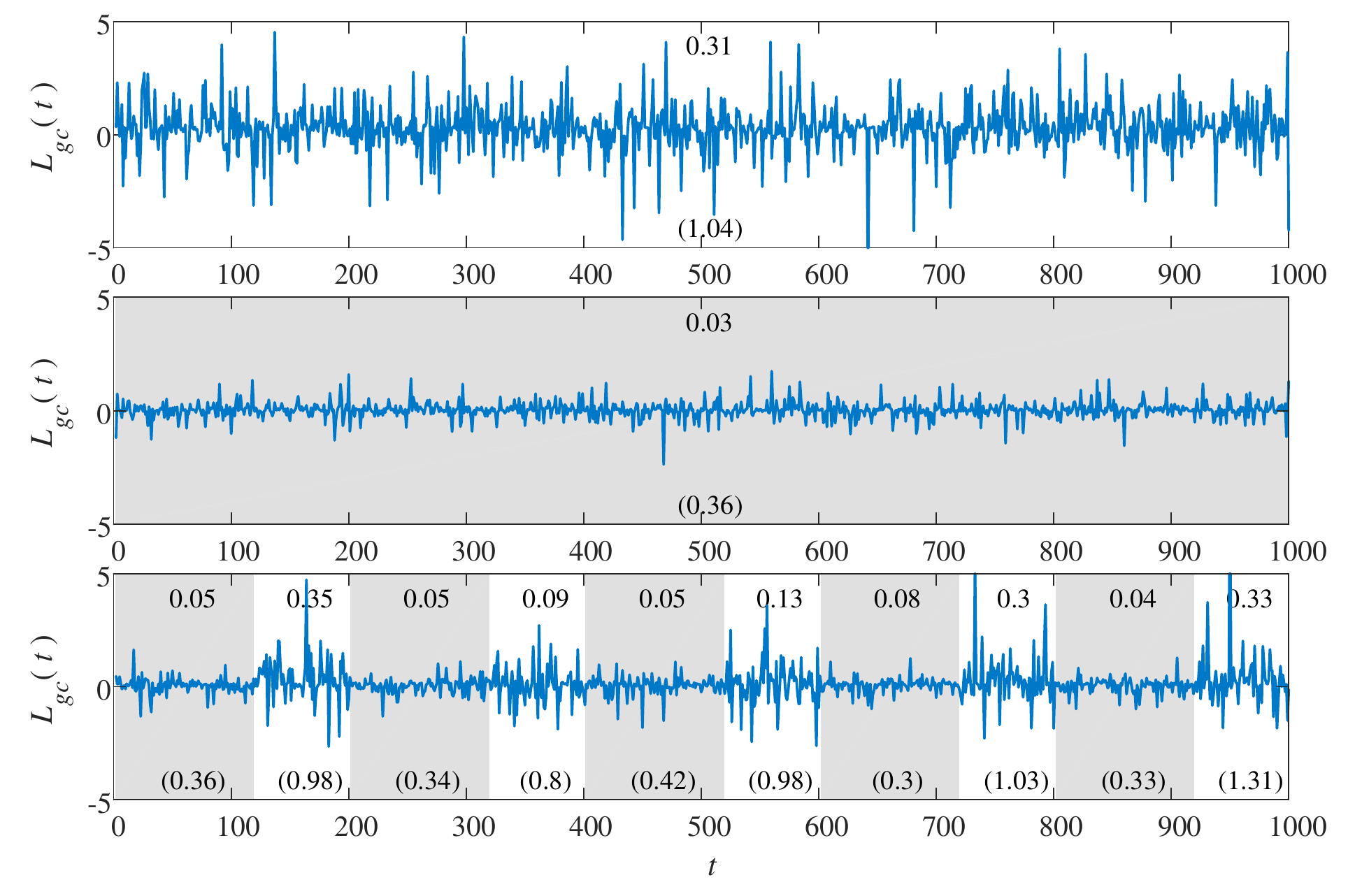}
\label{fig2}
\caption{Time course of the local GC computed for the toy model (\ref{eq:8}) along the direction $y \rightarrow x$  under different values of the standard deviation of the driving variable $y$, i.e. $\sigma^2_{\tilde{\epsilon}}=2$ (top), $\sigma^2_{\tilde{\epsilon}}=0.2$ (middle), and $\sigma^2_{\tilde{\epsilon}}$ alternating between 0.2 and 2 at the time points $t \in \{ 120, 200, 320, 400, 520, 600, 720, 800, 920\}$  (bottom).
The mean and standard deviation of the $L_{gc}$ computed within each time window where $\sigma^2_{\tilde{\epsilon}}=2$ (gray shaded epochs), or where $\sigma^2_{\tilde{\epsilon}}=0.2$ (white epochs), are reported respectively at the top and at the bottom (in brackets) of the window.}
\end{figure}

\begin{figure}[!ht] \centering
\includegraphics[width=15cm]{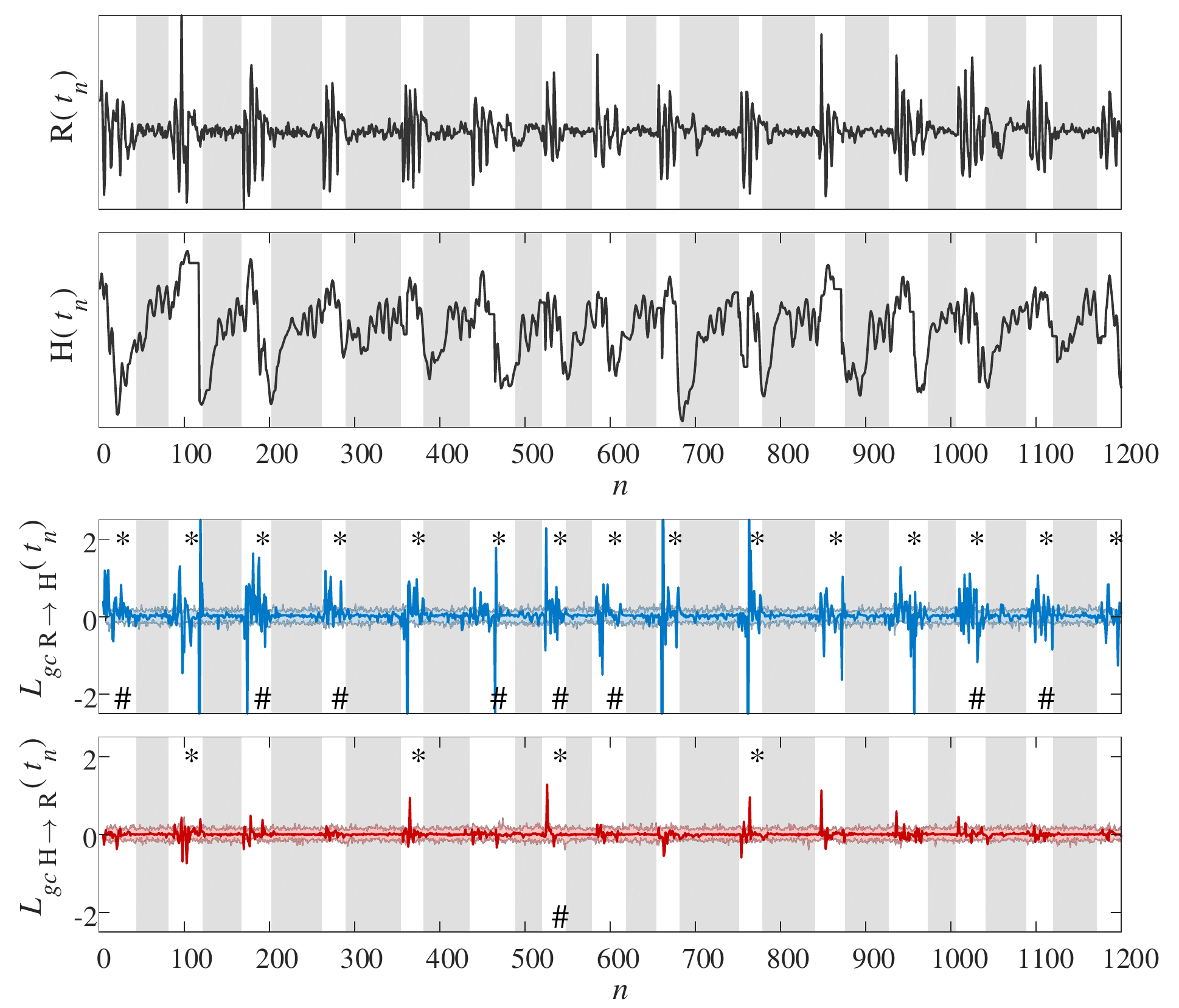}
\label{fig3}
\caption{Time series of respiration (R$(t_n)$) and heart rate (H$(t_n)$) measured for a subject exhibiting several instances of sleep apneas (gray shades in the plots), and corresponding time courses of the local GC computed from R to H ($L_{gc R \rightarrow H}(t_n)$) and from H to R ($L_{gc H \rightarrow R}(t_n)$). The discrete time points are $t_n=n\Delta T$, with $\Delta T=0.5 s$. The local GC courses are plotted together with the corresponding significance bounds (horizontal colored shades) obtained from IAAFT surrogates \cite{schreiberIAAFT}. The symbols $\#$ and $*$ mark statistically significant values of the mean and standard deviation of the local GC computed within each apneic or non-apneic time window (gray and white areas).}
\end{figure}

\begin{figure}[!ht] \centering
\includegraphics[width=15cm]{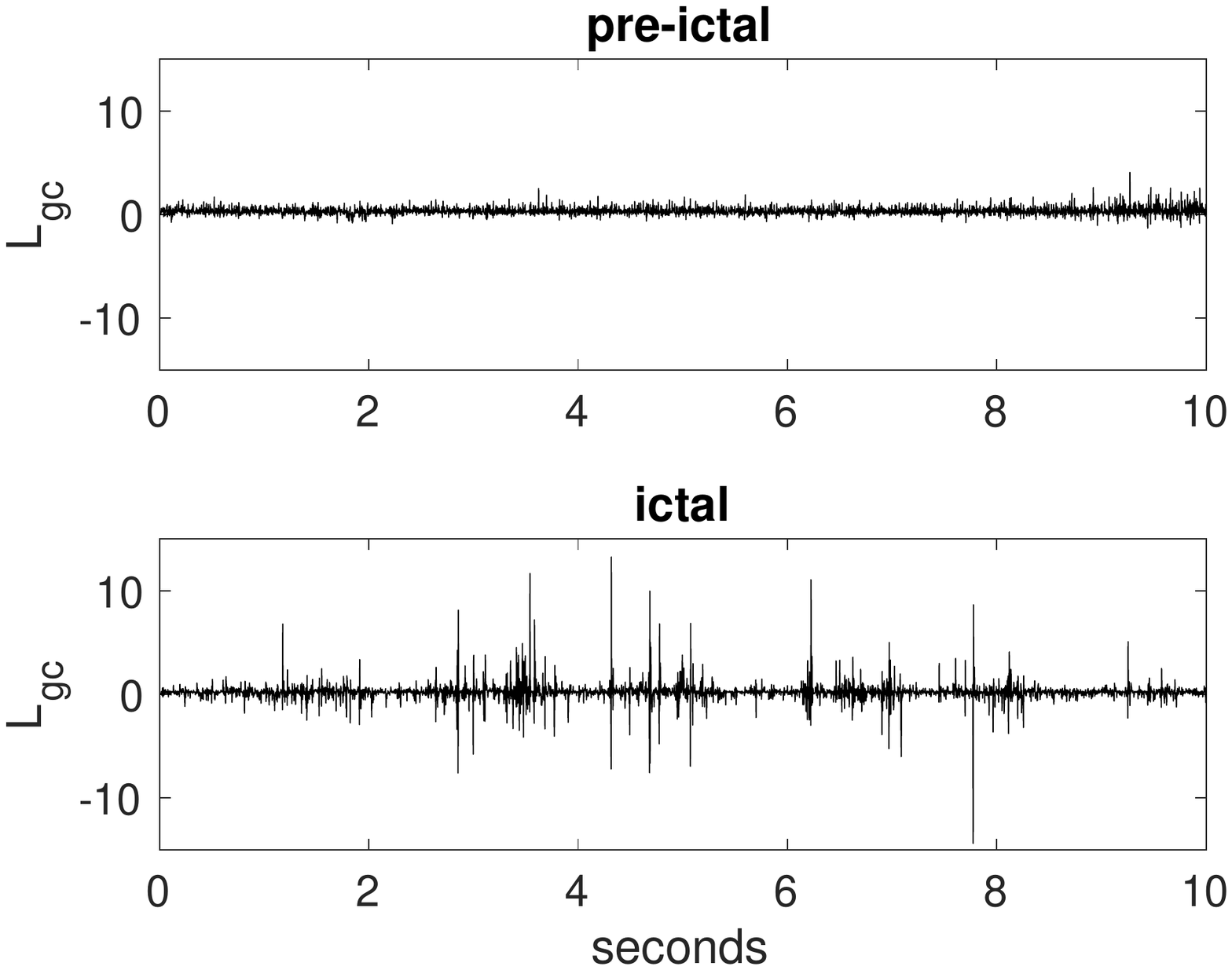}
\label{fig4}
\caption{Intracranial EEG in epilepsy: the average of local Granger causality from the depth electrode to the cortical target electrodes is depicted in the pre-ictal  and in the ictal stage.}
\end{figure}

\end{document}